# Reply to Werner

First, let me express me appreciation to Prof. Werner for his discussion of my paper. These are quite contentious and fiercely disputed questions, and the only way to make progress in resolving these disputes is to make a good faith effort to present the arguments clearly and explicitly. Only once things are pinned down on paper with some precision can a proper analysis proceed. Werner's willingness to do this means progress can be made.

There are some important points on which Werner and I agree. We agree that neither Einstein nor Bell presuppose determinism in framing their arguments, so the conclusions of those arguments cannot be avoided simply by adopting indeterminism. We even largely agree on what Bell *thought* he proved. For example, we agree that Bell did not take his theorem to refute determinism or "hidden variables". Werner even seems to concede that Bell himself thought that he had proven the necessity for any theory that reproduces the predictions of the quantum formalism—or more generally any theory that predicts violations of his inequality for pairs of experiments done at space-like separation—to be non-local (in a particular sense of "non-local"). Agreement on this much is already a tremendous amount of progress, and I do not wish to undervalue it.

But the remaining points of disagreement overwhelm this agreement in significance. Werner has made quite clear and explicit the startling claim that *Bell himself did not understand what he had proved*. If so, then Bell's own pronouncements about what he did, and what it means, are not reliable. Werner thinks that Bell and Einstein and I have all tacitly made an assumption of which we are unaware, an assumption he labels C for "classicality". When Bell, or Einstein, or I write "theory" what we really mean (although we don't realize it) is "*classical theory*". And when we draw conclusions about what a theory with certain characteristics must be like, the conclusions really only hold for *classical* theories. Furthermore, Operational quantum theory is not a classical theory. Therefore, according to Werner, Bell's and Einstein's conclusions simply do not apply to Operational quantum theory. In particular, Operational quantum theory can be local

*in Bell's and Einstein's sense* and still violate Bell's inequality because it is not classical. Werner concedes that Bell proved that any classical theory that violates his inequalities must be non-local (again, in Bell's and Einstein's sense of "non-local"). But deny classicality and the arguments no longer go through.

Bell and Einstein were human, and subject to error like all human beings. We should not dismiss out of hand the claim that Bell did not understand the significance of his own work, and that the major progress that has been made in the half-century since the publication of his landmark paper has been a proper understanding of what the paper proves, an understanding that eluded Bell himself. But by the same token, Werner and his colleagues are all human, and equally subject to error. *Someone* is making a mistake here. It could be Einstein and Bell (and me), it could Werner and those who think that Operational quantum theory falls outside the scope of Bell's conclusions. Fortunately, since we actually have the arguments down on paper, we can decisively determine that certain mistakes have been made. But before turning to that, let's briefly discuss the condition C.

The condition C is easily stated: it is that the state space of a theory forms a simplex. Good. The space of density matrices in quantum theory does not form a simplex, so if one takes the possible physical states of a system to be given by the density matrices, then one's theory is not classical in this sense. That much is clear. But what is not at all clear is *where the assumption that the state space is a simplex is presupposed in either Einstein's or Bell's reasoning*. One can search Werner's paper high and low for this vital piece of information and it is simply nowhere to be found. Which step of that argument, exactly, does not go through if the state space of the theory is not a simplex? We are given not a shred of an indication. This is a truly remarkable circumstance. Werner went to some pains in formulating his reply. He even tracked down the text of a letter by Born left out in a quotation of Bell's. Since the main contention is that Bell and Einstein and I have all been blinded by tacitly presuming classicality, the main order of business ought to be demonstrating *exactly where the argument presumes classicality*. But about this key, central question there is literally not a word, not a breath, not a clue.

For example, regarding the EPR argument Werner states that Einstein's "elements of reality" are "clearly intended as classical in the sense of C". I have given a rather painstaking and exact reconstruction of the EPR argument. Where in the course of that argument does any claim about the geometry of the state space of the theory play any role at all? If it does, the place ought to be easy to point out. And if there is no such place, then one can't avoid Einstein's conclusion simply by remarking that one's theory does not satisfy C. If Werner had done us the favor of actually explaining where C is presumed in the argument I could respond in more detail to his claim. Since he hasn't, I can do no better than assert: the geometry of the state space of a theory plays exactly no role at all in the EPR argument. The argument goes through no matter what the geometry of the state space is, whether it be a simplex or not.

Recall the dilemma posed by the EPR argument: if a theory predicts perfect correlations for the outcomes of distant experiments, then *either* the theory must treat these outcomes as deterministically produced from the prior states of the individual systems *or* the theory must violate EPR-locality. The argument is extremely simple and straightforward. The perfect correlations mean that one can come to make predictions with certainty about how system $S_1$ will behave on the basis of observing how the other, distant, system $S_2$ behaves. Either those observations of $S_2$ disturbed the physical state of $S_1$ or they did not. If they did, then that violates EPR-locality. If they did not, then $S_1$ must have been physically determined in how it would behave all along. That's the argument, from beginning to end. (That's also the point of Bell's discussion of Bertlmann's socks.) So preserving EPR-locality in these circumstances requires adopting a deterministic theory. Where, in this argument, does any presupposition about the geometry of the state space play any role? Nowhere.

But if Werner cannot actually identify any place where C is presupposed in Einstein's or Bell's reasoning, what makes him so sure that it is presupposed at all? It is here that we can really make solid progress. Werner is convinced that Bell and Einstein made a mistake not because he can actually identify any mistake but because he thinks that Operational quantum physics constitutes a *counterexample* to the conclusion Bell (and I) assert. That is, Werner thinks that Operational quantum physics is a theory that is EPR-local and also predicts violations of Bell's inequality for experiments done at space-like separation. If Operational quantum physics really were such a counterexample, then Bell (and I) have gone wrong somewhere. There would have to be some other presupposition to Bell's argument in addition to EPR-locality. It does not yet follow that C is the offending presupposition, but a counterexample is a counterexample. If, that is, Operational quantum theory *is* a counterexample.

Werner's claim here is made so clearly and so compactly that it is best to cite the entire passage, without ellipsis, to make sure nothing is left out:

> Consider this definition given in Maudlin's paper:
>> A physical theory is *EPR-local* iff_ according to the theory procedures carried out in one region do not immediately disturb the physical state of systems in sufficiently distant regions in any significant way.
>
> (Operational) quantum theory passes this criterion with flying colors. We can even drop the "immediately", the "sufficiently distant" the "in any significant way", and replace them by the absence of an explicit interaction between the subsystems. If there is an interaction, then of course there is an influence, which we can estimate in terms of the strength of the interaction and the duration it is turned on. Naturally, I have taken "physical state" here in the sense of the operational approach, as the quantity which allows us to determine the probabilities for all subsequent operations and measurements ("epistemic" rather than "ontic").

Note, in particular, the last sentence: Werner takes the physical state of a system to be it's *epistemic* state, the state on the basis of which we make actual predictions. Focus all of your attention on that sentence and on the definition of EPR-locality. Now one thing that is for sure, one thing that is beyond all possible dispute, is that the *epistemic* state assigned to $S_1$, the state we use to make *predictions* about how $S_1$ will behave, *is* changed as a consequence of observations made on $S_2$. Take a classic

EPR spin set-up: a pair of electrons is created in a singlet state and allowed to separate. One is sent to Alice, whose lab (with the electron) we will call $S_1$, and the other is sent to Bob whose lab is $S_2$. Before any operation is carried out in either lab, our *predictions* for the outcome of any spin measurement in either lab is 50% chance of "up" and 50% chance of "down". Suppose we know that Alice will measure spin in the *z*-direction. Can we *change* our epistemic state with regard to $S_1$ by carrying out procedures in $S_2$? Of course we can! In fact, by having Bob measure the *z*-spin of his electron, we can go from 50% confidence in our predictions for $S_1$ to complete certainty with respect to our prediction for $S_1$. *That's the whole point of the EPR argument*. The *epistemic* state assigned to $S_1$ certainly is changed, and disturbed, by the procedures carried out in $S_2$. The experiment carried out in $S_2$ provides information about how the experiment carried out in $S_1$ will come out. So *if one identifies the physical state with the epistemic state, as Werner explicitly and clearly does*, *the theory obviously fails to be EPR-local according to the cited definition*. The claim that Operational quantum physics "passes this criterion with flying colors" is pure bluff and manifestly false: having identified the physical and epistemic states, Operational quantum mechanics *fails* the criterion with *dive-bombing* colors. It fails already at the point of the EPR argument: violations of Bell's inequality do not even enter in.

Indeed, Werner's conceit that Operational quantum physics is some new-fangled theory, the likes of which never occurred to Einstein and Bell (and me) is a complete fabrication: Operational quantum physics is just plain-old vanilla Copenhagen quantum physics, the very theory that Einstein derided for its spooky action-at-a-distance. He derided it for exactly the reason illustrated in Werner's own presentation: by taking the *physical* state just to be the *epistemic* state, the theory already commits itself to violating EPR-locality in an EPR situation. The *predictive* state ascribed to $S_1$ is changed after observing the distant system $S_2$. So if the predictive state is the physical state, then the physical state changes. The geometry of the state space plays *no role at all* in this argument.

In fact, Werner has done us the great service of explicitly and clearly illustrating my main thesis: he fails to understand Bell because he has already failed to understand EPR. The passage cited above is proof. What is amazing is that the key sentence demonstrating the EPR-*non*-locality of Operational quantum theory comes only two sentences after the triumphant claim of its EPR-locality.

The only plausible explanation of this striking fact is that Werner cannot keep the various definitions of "locality" straight. Quantum theory is, of course, *signal*-local: one cannot use the physics to send superluminal messages. That is what is built into the Equal Time Commutation Relations of quantum field theory, for example. But signal-locality was never the issue for either Einstein or for Bell. If Einstein thought that standard quantum theory violated *signal*-locality then he would have proposed trying to build superluminal signaling devices. And Bell, of course, himself proves the "no Bell telephone" theorem, ruling out the possibility of superluminal signaling. So when Einstein complains that orthodox quantum theory, which *is* just Operational quantum theory, contains spooky action-at-a-distance he never meant to imply it contains the capacity to send superluminal signals. The only explanation I can come up with for the manifest incoherence of Werner's claims is that he thinks the signal-locality of quantum theory implies its EPR-locality. But that is a mistaken inference, as his own exposition of Operational quantum theory shows.

The EPR argument alone does not prove that the physical world fails to be EPR-local. That is what so frustrated Einstein: the *manifest* non-locality of Operational quantum theory appeared to him to be completely gratuitous. One can build an EPR-local theory that recovers the EPR correlations, but only by admitting determinism. For, as Bell stated, "any residual undeterminism could only spoil the perfect correlation". Einstein thought it was a blind commitment to *indeterminism* that was forcing the theory to be EPR-non-local even though the phenomena could be accounted for without by an EPR-local physics. It was Bell who proved Einstein wrong: EPR-locality cannot be reconciled with the full predictions of quantum theory after all. Certainly not by Operational quantum theory, as we have just seen. Werner's own characterization of the theory is as clear a proof of this as one could ask for. I hope that this exchange helps to clarify the situation.

One Last Remark

The referees for my paper expressed a wish that the paper discuss at least a bit what has happened in the 50 years since Bell's landmark paper rather just what happened in the 29 years preceding it. I hope that the central importance of understanding the EPR paper is now manifest, but a short comment about some more recent events is in order. One effect of not understanding what Bell did is that if his result is rederived in an unfamiliar way it can strike one as a momentous new discovery (which, indeed, it would be if it were new!). Just this has happened with John Conway and Simon Kochen's "Free Will Theorem". The theorem is presented in an unusual garb, dressed in the language of free will. Even more striking, the following claim is made: "if indeed we humans have free will, then elementary particles already have their own small share of this valuable commodity" [1]! At first glance, this seem miles away from Bell's result.

But properly translated, it is just a corollary of Bell's theorem. "Humans having free will" just means that operations under experimental control, such as picking which "measurement" to perform, are treated as free variables. More precisely, the choice of experimental arrangement is taken to be statistically independent of the initial state of the particles. This is the rejection of what has been called "superdeterminism", and is also used in Bell's derivation. And "elementary particles having free will" just means that the outcome of an experiment on a particle is not determined by the physical state in the particle's backward light-cone. If the outcome were always so determined, then the theory would be EPR-local. So properly translated, the "Free Will Theorem" just says that barring superdeterminism (and granting quantum-mechanical predictions), a theory cannot be EPR-local. Just as Bell said. For an analysis of the Free Will theorem in much more exacting detail, see [GTTZ10].

It is not at all surprising that Conway and Kochen should be impressed by this theorem: it truly is an amazing result! It just happens to be what Bell had already proved. That's why we still have to attend carefully to what Bell did. If it remains unappreciated it is bound to be rediscovered, but without the realization that the result, while astonishing, is not new.

References

[1] Conway, J. and Kochen S. 2009 *Notices of the American Mathematical Society* **56** (2), 226-232

[2] Goldstein, S., Tausk, D., Tumulka, R. and Zanghì, N. 2010 *Notices of the American Mathematical Society* **57** (11), 1451-1453